\documentclass{ws-ijmpd}
\usepackage[super,compress]{cite}
\usepackage{url}
\usepackage[breaklinks]{hyperref}
\hypersetup{colorlinks,urlcolor=black,citecolor=black,linkcolor=black,filecolor=black}

\newcommand{\beq}{\begin{equation}}
\newcommand{\eeq}{\end{equation}}
\newcommand{\beqa}{\begin{eqnarray}}
\newcommand{\eeqa}{\end{eqnarray}}

\newcommand{\diff}{\mathrm{d}}

\usepackage{graphicx}

\begin{document}

\markboth{Rafael P. Bernar}
{Scalar radiation emitted by a source around a spherically symmetric BH}

%
\catchline{}{}{}{}{}
%

\title{Scalar radiation emitted by a source around a spherically symmetric black hole}

\author{Rafael P. Bernar}
\address{{Faculdade de F\'{\i}sica, Universidade 
Federal do Par\'a, 66075-110, Bel\'em, Par\'a, Brazil}\\
rafael.bernar@icen.ufpa.br\\}

\maketitle

\begin{history}
\received{Day Month Year}
\revised{Day Month Year}
\end{history}

\begin{abstract}
We analyze the scalar radiation emitted by a source in a circular geodesic orbit around a spherically symmetric black hole. The black hole spacetime considered is quite general, in the sense that it encompasses the solutions of Schwarzschild and Reissner-Nordstr\"{o}m, and also the Bardeen solution of a regular black hole. We use the framework of quantum field theory in curved spaces to compute the one-particle emission amplitude of scalar particles and related quantities.  
\end{abstract}

\keywords{gravitational waves; static spacetime; circular motion.}

\ccode{PACS numbers: 04.60.-m, 
	04.62.+v, 
	04.50.-h, 
	04.25.Nx, 
	04.60.Gw, 
	11.25.Db  
}


\section{Introduction}	

Black holes (BHs) are one of the most important predictions of General Relativity (GR). Their strong gravity regime and nontrivial causal structure provide a very rich setting for the study of a plethora of different phenomena. Moreover, BH spacetimes are central objects in the discussion of quantum theories of gravitation, where they are believed to play an important part. Understanding BH interactions with fundamental fields is, therefore, of paramount importance. 

By looking at the effect that the spacetime curvature has on the field observables, such as absorption and scattering data\cite{Matzner1968,Sanchez1978,Macedo2013} or the field quasinormal modes,\cite{Nollert1999,Berti2009,Konoplya2011} we may infer some of the BH properties. In particular, the study of test fields through radiation emission by matter in the vicinity of BHs provides insights both to how matter behaves in strong gravity regimes and to the dynamics of spacetime itself. Additionally, the recent detection of gravitational waves radiated by a BH binary merger\cite{gw150914} have intensified efforts in related research subjects. 

According to the weak cosmic censorship conjecture,\cite{Penrose1969,Wald1997} singularities in GR are protected by event horizons, what means that they are located inside an event horizon. Moreover, the singularity theorems impose very restrictive constraints on the types of matter that can avoid their formation through gravitational collapse.\cite{Hawking1975a} Regular BHs, those with no singularities, are solutions of the Einstein equations, but, to circumvent the singularity theorems in GR, the gravitating source usually violates some energy condition. Thus, such a source must consist of exotic matter. The first known regular BH is the Bardeen solution.\cite{Bardeen1968} Nevertheless, regular BH spacetimes may also appear as an effective spacetime description in a quantum theory of gravity.\cite{Carballo-Rubio2018} The Bardeen BH scalar absorption and scattering spectra were analyzed in Refs.~\refcite{Macedo2014} and \refcite{Macedo2015}. The quasinormal modes were also computed.\cite{Flachi2012,Macedo2016} However, the study of radiation emission in the vicinity of these objects has received less attention in the literature.

The scalar radiation emitted by a source in circular geodesic orbit around a Schwarzschild BH was analyzed in Refs.~\refcite{misnergsr} and \refcite{Breuer1973}. In this case, the source can emit synchrotron radiation, that is, radiation distributed in narrow angles, mainly concentrated in the plane of orbit, and with frequencies much higher than the angular frequency of the orbit. The scalar field was used as a first approximation model to more realistic settings. Nevertheless, this simple model can qualitatively capture several features of the electromagnetic or gravitational radiation cases. Using a quantum field theory in curved spacetime approach, several radiation settings near BHs were studied, where the emitted radiation is investigated by computing the one-particle emission amplitude at tree level, due to the field being excited by an external source. For a source rotating around a Schwarzschild BH, the scalar,\cite{Crispino2000,Castineiras2007,Crispino2008} electromagnetic\cite{Castineiras2005} and gravitational\cite{Bernar2017} radiation have been analyzed. It was found that, although quite similar to the scalar and electromagnetic radiations, the polarization present in the gravitational case presents an obstacle for synchrotron radiation to occur, in accordance with past results.\cite{Davis1972,Breuer1973a} For rotating BHs, the scalar radiation case was computed in Ref.~\refcite{Macedo2012}. There are also results for the Reissner-Nordstr\"{o}m spacetime,\cite{Crispino2009} as well as for Schwarzschild-anti-de Sitter geometry.\cite{Cardoso2002} Although equivalent to the classical results, this semiclassical approach naturally extends to fields of different spins. Additionally, the introduction of quantum effects, such as radiative corrections or the interaction with quantum sources,\footnote{A two-level Unruh detector, for example.} is readily available with this method.

The purpose of this paper is to extend the works in Refs.~\refcite{Crispino2000}, \refcite{Crispino2008}, \refcite{Crispino2009} and \refcite{Bernar2019}, by considering a general spherically symmetric BH spacetime, instead of a specific BH case. We also compute the radiation spectral distribution, displaying its synchrotron behavior.
In Sec.~\ref{sec:scalarfield}, we give some details of the spacetime of a spherically symmetric BH and present the quantization of a massless scalar field in this curved spacetime. The scalar source and emitted radiation are presented in Sec.~\ref{sec:scalar-source-radiation}. We present some of our results in Sec.~\ref{sec:results}. We conclude with a discussion in Sec.~\ref{sec:conclusion}. Throughout, we use natural units ($c=G=\hbar=1$), and the metric signature ($-,+,+,+$).

\section{Quantization of the Scalar Field in Spherically Symmetric Spacetimes}
\label{sec:scalarfield}

We assume the BH spacetime to be asymptotically flat, spherically symmetric and static. Additionally, it is regular everywhere except at the origin, where it can be singular. The line element may then be written as 
\beq
\diff s^2=-f(r) \diff t^2+g(r)\diff r^2+r^2(\diff \theta^2+\sin \theta^2 \diff \phi^2), \label{eq:sphericalmetric}
\eeq
where $(t,r,\theta,\phi)$ are Schwarzschild-like coordinates. The event horizon location, $r_{+}$, is given by solving 
\beq
f(r_{+})=0.
\eeq
The line element given by Eq.~(\ref{eq:sphericalmetric}) describes a static, electrically charged BH, a solution of the Einstein-Maxwell system, if the functions $f$ and $g$ are given by
\beq
f(r)=g(r)^{-1}=1-\frac{2M}{r}+\frac{q_{RN}^2}{r^2}, \label{eq:RN-function}
\eeq
where $M$ is the BH mass, $q_{RN}$ is the electric charge and we have that $0 \leq|q_{RN}| \leq M$. This is the so-called Reissner-Nordstr\"{o}m solution. The event horizon is located at
\beq
r_{+}=M+\sqrt{M^2-q_{RN}^2}.
\eeq
Note that the Schwarzschild solution is recovered when $q_{RN}=0$, with the event horizon located at $r_{+}=2M$.

The spacetime of a regular BH, the Bardeen solution,\cite{Bardeen1968} can also be associated to the line element given by Eq.~(\ref{eq:sphericalmetric}), with
\beq
f(r)=g(r)^{-1}=1-\frac{2 M r^2}{(r^2+q_B^2)^{\frac{3}{2}}}.
\eeq
The parameter $M$ is associated to the BH mass, while the parameter $q_B$ is the charge of a magnetic monopole in a theory of nonlinear electrodynamics.\cite{Ayon-Beato2000} This solution can also be associated to an electric source, in a different nonlinear electrodynamics theory.\cite{Rodrigues2018} In the context of nonlinear electrodynamics, other electrically charged regular BH solutions have also been found.\cite{Beato1998,Balart2014} 
An event horizon is located at $r_{+}$ satisfying $f(r_{+})=0$, provided that $0 \leq |q_B| \leq 4 M/(3 \sqrt{3})$. There is no singularity at the origin $r=0$ and the function $f(r) \approx 1-2Mr^{2}/q_B^3$ behaves similar to the one found in de Sitter space in the static patch. This solution violates the strong energy condition inside the BH.\cite{Zaslavskii2010} In order to compare the Bardeen BH case with the RN BH case, we define a normalized charge, with which we parametrize our results, namely
\beq
Q_{(i)} \equiv q_{(i)}/q^{\mathrm{ext}}_{(i)},
\eeq
with $(i)=B, RN$, $q^{\mathrm{ext}}_{B}=4 M/(3 \sqrt{3})$ and $q^{\mathrm{ext}}_{RN}=M$ are the charges of the extreme cases.

The quantization of the massless scalar field in a general spherically symmetric BH spacetime follows closely the quantization procedure in the spacetime of a Schwarzschild BH.~\cite{Crispino2000} The scalar field $\Phi(x)$ obeys the massless Klein-Gordon equation, given by  
\beq
\nabla^{\mu}\nabla_{\mu} \Phi(x) =0. \label{eq:scalarfieldeq}
\eeq

\begin{figure}[!th]
	\centering
	\includegraphics[scale=1]{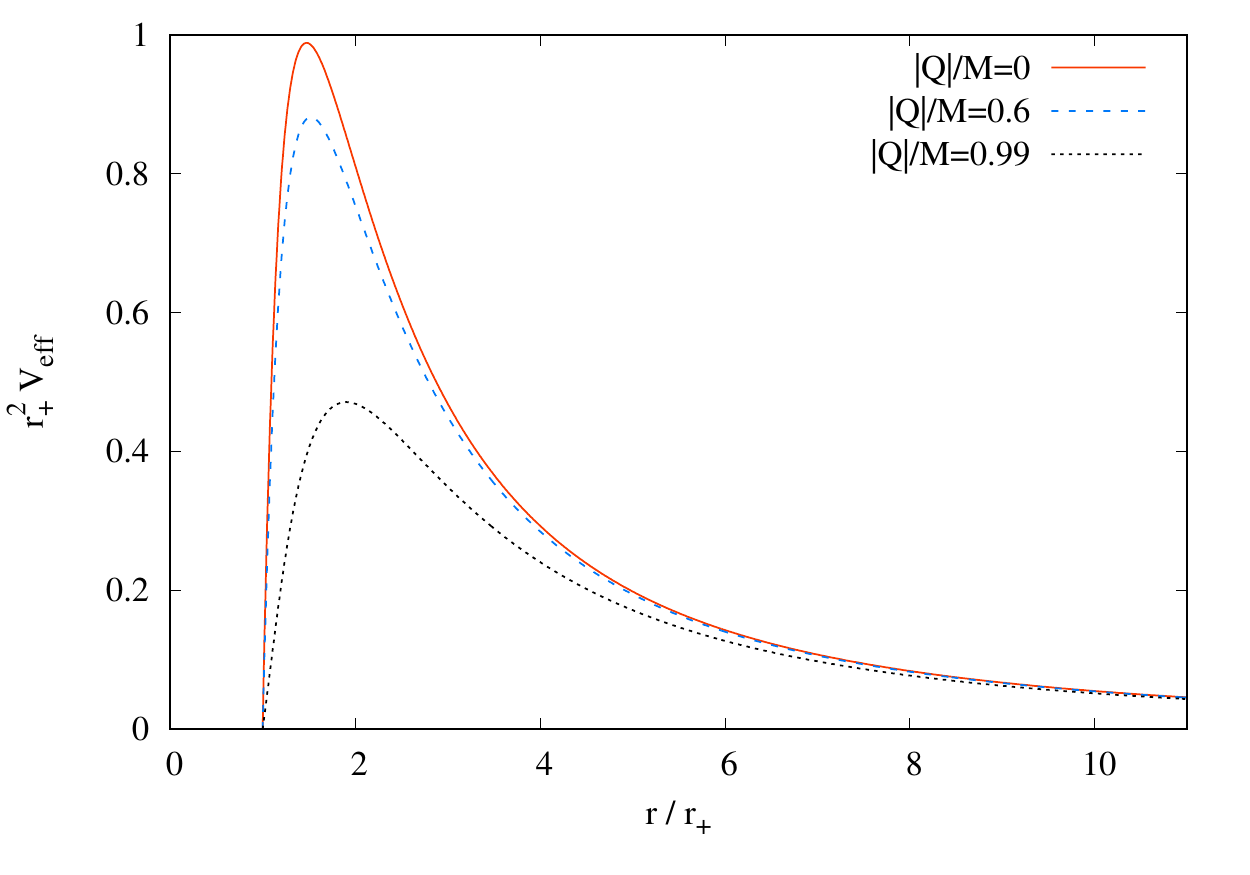}
	\caption{Effective potential, given by Eq.~(\ref{eq:effective-potential}), plotted for $l=2$, as a function of $r/r_{+}$ for the Schwarzschild BH case ($|Q_B|=0$) and the Bardeen BH cases with the $|Q_B|=0.6$ and $|Q_B|=0.99$.}
	\label{fig:effectivepontential}
\end{figure}

A possible basis of solutions to Eq.~(\ref{eq:scalarfieldeq}), which are of positive frequency with respect to the timelike Killing vector field $\partial_t$, may be given as
\beq
u_{\omega l m}(x)=\sqrt{\frac{\omega}{\pi}}\frac{\psi_{\omega l}(r)}{r}Y_{lm}(\theta,\phi)e^{-i \omega t} \ \ \ (\omega > 0), \label{eq:field-decompostion}
\eeq
where $Y_{lm}(\theta,\phi)$ are the scalar spherical harmonics, normalized such that
\beq
\int_{S^2}\sin \theta \diff\theta \diff \phi \overline{Y_{lm}}Y_{l'm'}=\delta_{ll'}\delta_{mm'}.
\eeq 
The overbar denotes complex conjugation.
The functions $\psi_{\omega l}(r)$ satisfy the equation:
\beq
\left[-\sqrt{\frac{f(r)}{g(r)}}\frac{\diff}{\diff r}\left(\sqrt{\frac{f(r)}{g(r)}}\frac{\diff}{\diff r}\right)+V_{\mathrm{eff}}(r)\right]\psi_{\omega l}(r)=\omega^2\psi_{\omega l}(r), \label{eq:radial-part-eq}
\eeq
where the effective potential is written as (see Fig.~\ref{fig:effectivepontential})
\beq
V_{\mathrm{eff}}(r)=f(r)\left\{\frac{l(l+1)}{r^2}+\frac{1}{2 r f(r)}\frac{\diff}{\diff r}\left[\frac{f(r)}{g(r)}\right]\right\}. \label{eq:effective-potential}
\eeq

\begin{figure}[!h]
	\centering
	\includegraphics[scale=1]{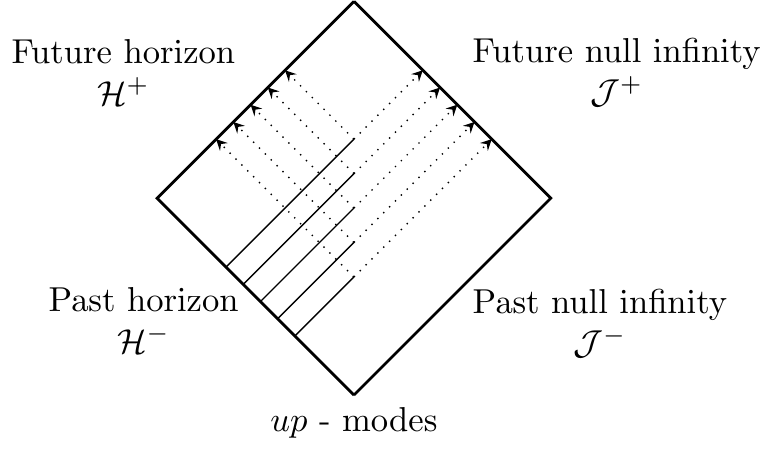}
	\caption{The $up$ modes in the Penrose diagram of the BH outside region.}
	\label{fig:penrosediagram-up}
\end{figure}

\begin{figure}[!h]
	\centering
	\includegraphics[scale=1]{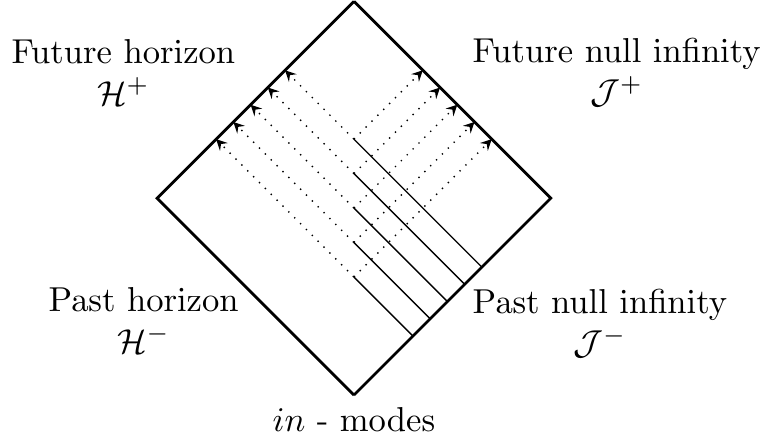}
	\caption{The $in$ modes in the Penrose diagram of the BH outside region.}
	\label{fig:penrosediagram-in}
\end{figure}

We choose the two independent solutions to Eq.~(\ref{eq:radial-part-eq}) to have the following asymptotic forms
\beq
\psi^{up}_{\omega l}\approx \begin{cases} A^{up}_{\omega l}\left(e^{i\omega r^{*}}+\mathcal{R}^{up}_{\omega l}e^{-i\omega r^{*}}\right), \hspace{0.25cm}r \gtrsim r_{+},\\
	A^{up}_{\omega l}\mathcal{T}^{up}_{\omega l}e^{i\omega r^{*}}, \hspace{0.5cm} r \gg r_{+}, \end{cases} \label{eq:up-boundary-conditions}
\eeq
\beq
\psi^{in}_{\omega l} \approx \begin{cases} A^{in}_{\omega l}\mathcal{T}^{in}_{\omega l}e^{-i\omega r^{*}}, \hspace{0.5cm}r \gtrsim r_{+},\\
	A^{in}_{\omega l}\left(e^{-i \omega r^{*}}+\mathcal{R}^{in}_{\omega l}e^{i \omega r^{*}}\right), \hspace{.25cm} r \gg r_{+}, \end{cases} \label{eq:in-boundary-conditions}.
\eeq
The tortoise coordinate, $r^{*}$, can be computed by 
\beq
\diff r^{*} \equiv \sqrt{\frac{g(r)}{f(r)}}\diff r.
\eeq
We assume the existence of a past event horizon, i.e. the BH is eternal. The $up$ solutions are associated with modes purely incoming from the past event horizon $\mathcal{H}^{-}$, while the $in$ solutions are associated with modes purely incoming from the past null infinity $\mathcal{J}^{-}$. This is pictorially represented in Figs.~\ref{fig:penrosediagram-up} and \ref{fig:penrosediagram-in}.

We expand the quantum field operator $\hat{\Phi}(x)$ as
\beq
\hat{\Phi}(x)=\sum_{P} \sum_{l,m}\int_{0}^{\infty} \diff \omega \left[ u^{P}_{\omega l m}(x) \hat{a}^{P}_{\omega l m} + \mathrm{H.c.} \right],
\eeq
where the superscript $P$ labels the $in$ and $up$ modes.
The modes $u^{P}_{\omega l m}(x)$ are normalized according to the Klein-Gordon inner product, defined by
\beq
\langle \Phi, \Psi \rangle \equiv i \int_{\Sigma} \diff \Sigma^{\mu} \left(\overline{\Phi}\nabla_{\mu}\Psi-\Psi\nabla_{\mu}\overline{\Phi}\right), \label{eq:innerproduct}
\eeq
where $\Sigma$ is a Cauchy hypersurface. The inner product (\ref{eq:innerproduct}) does not depend on the choice of the hypersurface $\Sigma$, if $\Phi$ and $\Psi$ satisfy the equations of motion.~\cite{Friedman1978} We require
\beq
\langle u^{P}_{\omega l m}, u^{P'}_{\omega' l' m'} \rangle=\delta^{PP'}\delta^{ll'}\delta^{mm'}\delta(\omega-\omega'),
\eeq
and we obtain the overall normalization constants in Eqs.~(\ref{eq:up-boundary-conditions}) and~(\ref{eq:in-boundary-conditions}), namely
\beq
A^{in}_{\omega l}=A^{up}_{\omega l}=\frac{1}{2 \omega}. \label{eq:normalizationconstants}
\eeq
To obtain the normalization given by Eq.~(\ref{eq:normalizationconstants}), we need to use Eq.~(\ref{eq:radial-part-eq}), that $\psi_{\omega l}$ (and $\overline{\psi_{\omega l}}$ as well) satisfy, together with
\beq
\frac{1}{\omega'-\omega}\left.\left[\psi^{P'}_{\omega'l}\frac{\diff \overline{\psi^{P}_{\omega l}}}{\diff r^{*}}-\overline{\psi^{P}_{\omega l}}\frac{\diff \psi^{P'}_{\omega' l}}{\diff r^{*}}\right]\right|^{r^{*} \to +\infty}_{r^{*} \to -\infty}=4 \pi \omega |A_{\omega l}|^2 \delta^{PP'}\delta(\omega - \omega').
\eeq
With the normalization above, the operators, $\hat{a}^{P\dagger}_{\omega l m}$ and $\hat{a}^{P}_{\omega l m}$, satisfy the nonvanishing commutation relations, given by
\beq
[\hat{a}^{P}_{\omega l m},\hat{a}^{P'\dagger}_{\omega' l' m'}]=\delta^{PP'}\delta^{ll'}\delta^{mm'}\delta(\omega-\omega')
\eeq
and have the usual interpretation of creation and annihilation operators, respectively.

\section{Scalar Radiation from a Source in Circular Geodesic Motion}
\label{sec:scalar-source-radiation}

Let us now compute the scalar radiation emitted by a source in circular geodesic orbit around the BH. We consider an external scalar source $j(x)$ interacting with the scalar field via the interaction action
\beq
\hat{S}_I=\int \diff^4 x \sqrt{-g} \ j(x)\hat{\Phi}(x).
\eeq
The source moves with constant angular velocity $\Omega$,\footnote{This quantity is measured by asymptotic static observers.} at $r=R$, in the plane $\theta=\pi/2$. Hence the source current is given by
\beq
j(x)=\frac{\lambda}{u^{t}\sqrt{-g}}\delta(r-R)\delta(\theta - \pi/2)\delta(\phi - \Omega t), \label{eq:source-current}
\eeq
where $\lambda$ is a small coupling constant. The $4$-velocity of the source is 
\beq
u^{\mu}=(\gamma,0,0,\gamma \Omega),
\eeq
where
\beq
\gamma=\frac{\diff t}{\diff \tau}=\frac{1}{[f(R)-R^2\Omega^2]^{\frac{1}{2}}}.
\eeq
The orbital angular velocity, as a function of the orbit radius $R$, is given by
\beq
\Omega(R)=\sqrt{\frac{f'(R)}{2 R}}. \label{eq:angular-velocity}
\eeq
We note in passing that $g(r)$ plays no role in the source's motion, since there is no change in its radial position. (For a review of geodesics in a general spherically symmetric spacetime, we refer the reader to Ref.~\refcite{Cardoso2009}.)

The one-particle emission amplitude may be computed to first order in perturbation theory and is given by
\beq
\mathcal{A}^{P;\omega l m}_{\mathrm{em}}= \langle P; \omega l m| i \hat{S}_I | 0 \rangle = i \int \diff^4 x \sqrt{-g} \ j(x) \overline{u^{P}_{\omega l m}}(x). \label{eq:emissionamplitude}
\eeq
This is the amplitude for emission of a $P=in,up$ scalar particle with a given frequency $\omega$ and angular quantum numbers $l$ and $m$. The vacuum state is the one defined by
\beq
\hat{a}^{P}_{\omega l m} |0 \rangle =0.
\eeq  
This vacuum state plays the same role as the Boulware vacuum in Schwarzschild spacetime.\cite{Boulware1975} If we had considered the Unruh-\cite{Unruh1976} or Hartle-Hawking-like vacuum states,\cite{Hartle1976} the power would be the one associated with the net emitted radiation.~\cite{Bernar2019} The emission amplitude, given by Eq.~(\ref{eq:emissionamplitude}), is proportional to
\beq
\int e^{i (\omega -m \Omega)t}\diff t=2\pi \delta(\omega -m \Omega),
\eeq
 i.e. there is only emission of scalar particles with $\omega=m \Omega$. In particular, only the modes with $m \geq 1$ contribute to the emitted radiation. 

The emitted power due to a given mode of a $P$-type scalar particle, with frequency $\omega$ and quantum numbers $l$ and $m$, is then given by
\beq
W^{P;lm}_{\mathrm{em}}=\int_{0}^{\infty} \diff \omega \ \omega \ |\mathcal{A}^{P;\omega l m}_{\mathrm{em}}|^2 / T,
\eeq
where $T=\int_{-\infty}^{\infty} \diff t=2 \pi \delta(0)$ is the total time that the source spents emitting radiation (as measured by an asymptotic static observer)~\cite{Higuchi1998,Crispino1998}. Using the source given by Eq.(~\ref{eq:source-current}), we obtain the following emitted power

\beqa
W^{P;lm}_{\mathrm{em}}=2 \omega_m^2 \lambda^2 [f(R)-R^2 \Omega^2]\left|\frac{\psi^{P}_{\omega_m l}}{R}\right|^2\left|Y_{lm}(\pi/2,\Omega t)\right|^2, \nonumber \\
\eeqa
where
\beq 
\omega_m \equiv m \Omega. \label{eq:frequency-m}
\eeq
The $\left|Y_{lm}(\pi/2,\Omega t)\right|^2$ is time independent, it vanishes for odd values of $l+m$ and 
\beq
\left|Y_{lm}(\pi/2,\Omega t)\right|^2=\frac{2l+1}{4 \pi}\frac{(l+m-1)!!(l-m-1)!!}{(l+m)!!(l-m)!!}
\eeq
for even values of $l+m$~\cite{Gradshteyn1980}. Hence, for a certain frequency $\omega_m=m \Omega$ of the emitted radiation, the multipole modes that contribute to the emitted power are the ones with $l=m,m+2,m+4,...$.

The total emitted power is given by
\beq
W_{\mathrm{em}}=\sum_{P=in, up}\sum_{l,m}W^{P;lm}_{\mathrm{em}}. \label{eq:total-emitted-power}
\eeq
To compute the amount of radiation that is observed at infinity, we have to consider the modes that are purely outgoing at future null infinity $\mathcal{J}^{+}$, i.e. no radiation entering the inside region of the BH. These are the $out$ modes, related to the $in$ modes by complex conjugation. Hence, the observed power at infinity is obtained by summing only the contributions from the $in$ modes in Eq.~(\ref{eq:total-emitted-power}), namely\cite{Crispino2016}
\beq
W^{in}_{\mathrm{em}}=\sum_{l,m}W^{in;lm}_{\mathrm{em}}, \label{eq:observed-power}
\eeq
which can be compared to the total emitted power.

\section{Results}
\label{sec:results}

We integrate Eq.~(\ref{eq:radial-part-eq}) to obtain the radial functions $\psi^{up}_{\omega l}$ and $\psi^{in}_{\omega l}$. For this, the same numerical procedure as the one in Ref.~\refcite{Bernar2019} is used. We invert Eq.~(\ref{eq:angular-velocity}) to obtain $R$ as a function of $\Omega$. In this manner, the emitted power is computed as a function of quantities measured at infinity ($M$ and $\Omega$).

\begin{figure}
	\centering
	\includegraphics[scale=1]{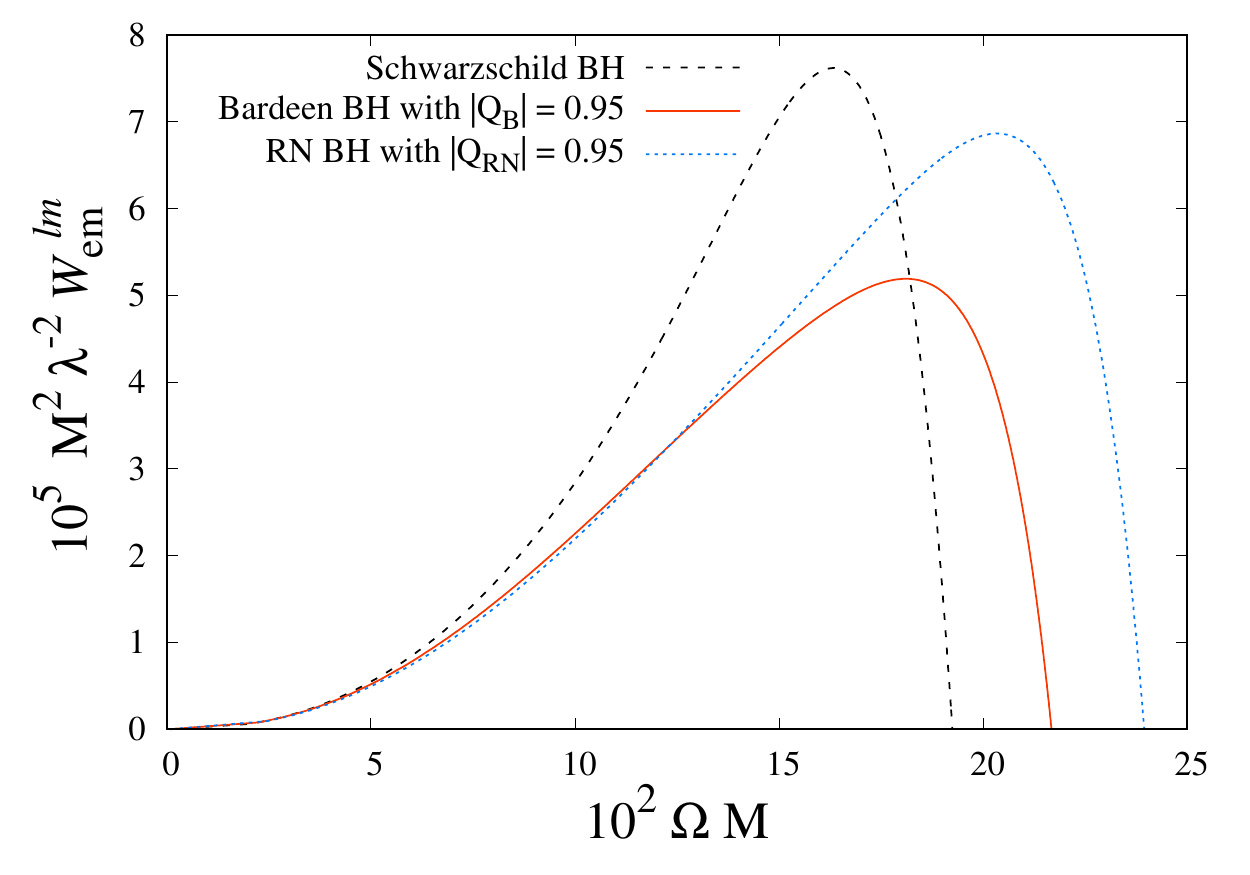}
	\caption{Emitted power of the mode $l=m=1$, as a function of $\Omega$, for the scalar source in circular orbit around the following BHs: (i) a Schwarzschild BH; (ii) a Bardeen BH with $|Q_B|=0.95$; (iii) a Reissner-Nordstr\"{o}m BH with $|Q_{RN}|=0.95$.}
	\label{fig:comparison-W}
\end{figure}

In Ref.~\refcite{Bernar2019}, some results for the scalar radiation from a source rotating around a Bardeen BH have been obtained. For completeness, we briefly describe some of them here. In comparison with the RN BH case with the same normalized charge, the emitted power, for a given choice of $l$ and $m$, namely
\beq
W^{lm}_{\mathrm{em}}=W^{in;lm}_{\mathrm{em}}+W^{up;lm}_{\mathrm{em}},
\eeq 
is greater in the Bardeen BH case for low angular velocities, as can be seen in Fig.~\ref{fig:comparison-W}. This is mainly due to the behavior of the metric far away from the BH (where $\Omega \to 0$). The charge contribution to the metric falls off faster in the Bardeen BH case than in the RN BH case. Thus, far away from the BH, that is, in the region where $\Omega \to 0$, the emitted power in the Bardeen case, being greater than the RN BH one,\footnote{Nevertheless, the RN BH and Bardeen BH emitted powers are still very similar for low angular velocities.} is closer to the Schwarzschild BH emitted power.  


\begin{figure*}[!h]
	\centering
	\includegraphics[scale=0.5]{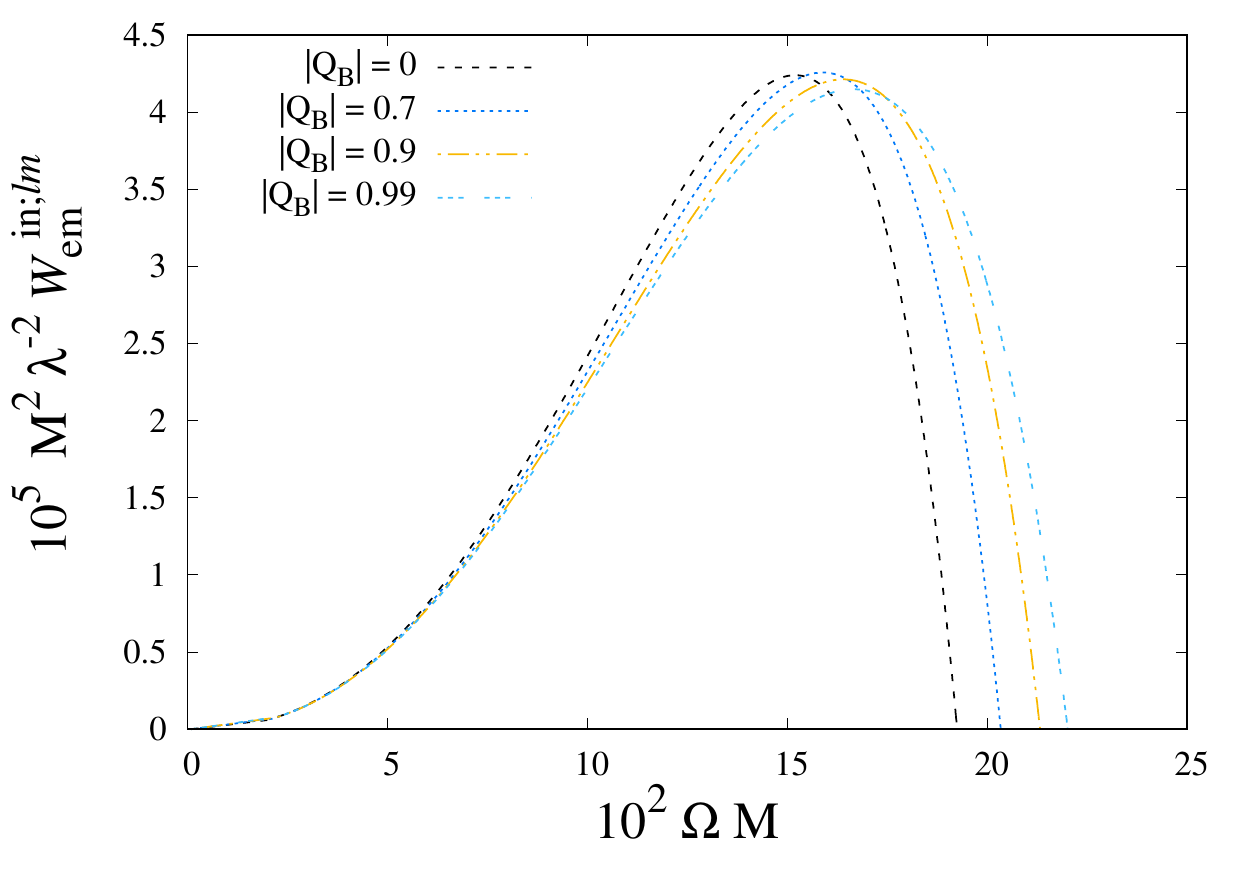} \includegraphics[scale=0.5, trim={1.9cm 0 0 0}, clip]{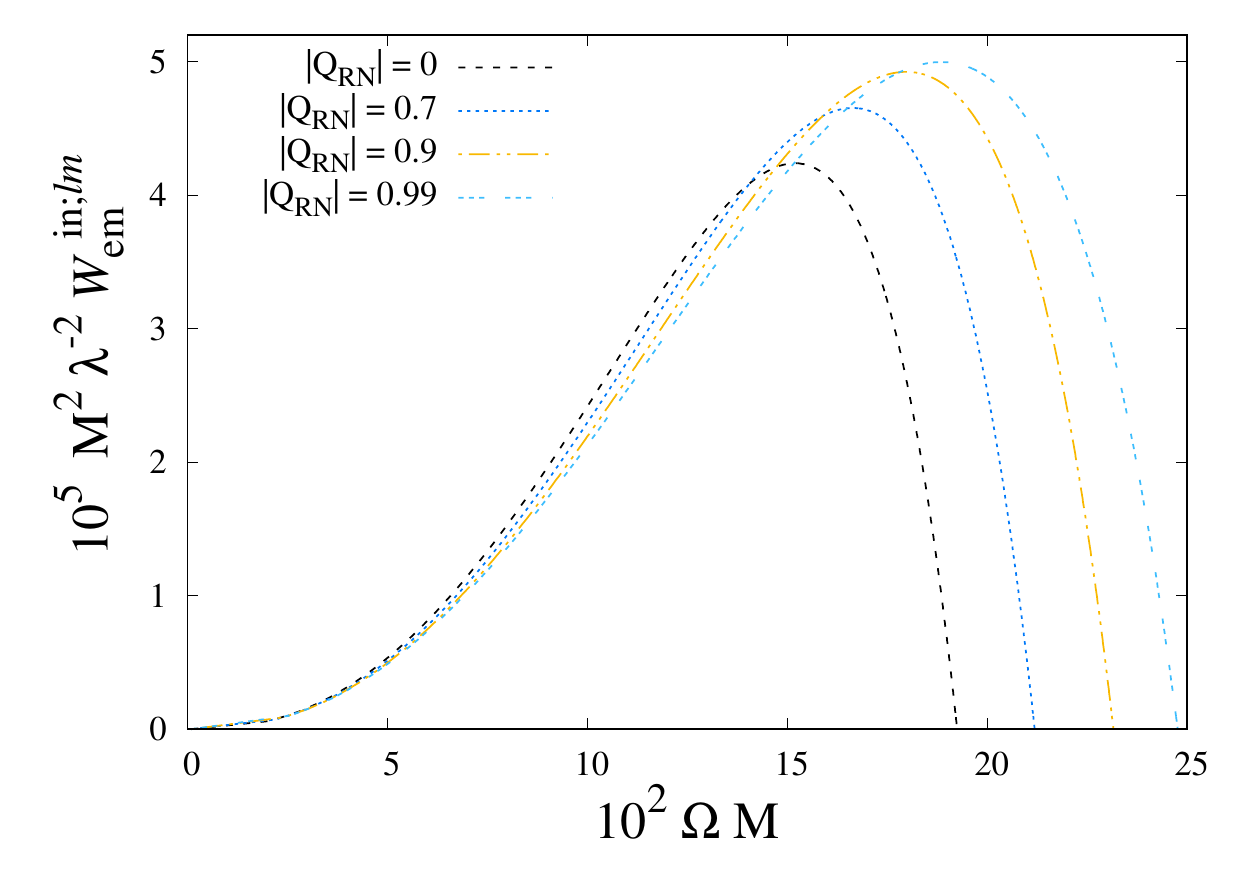}  \\
	\vspace{0.33cm}
	\includegraphics[scale=0.5]{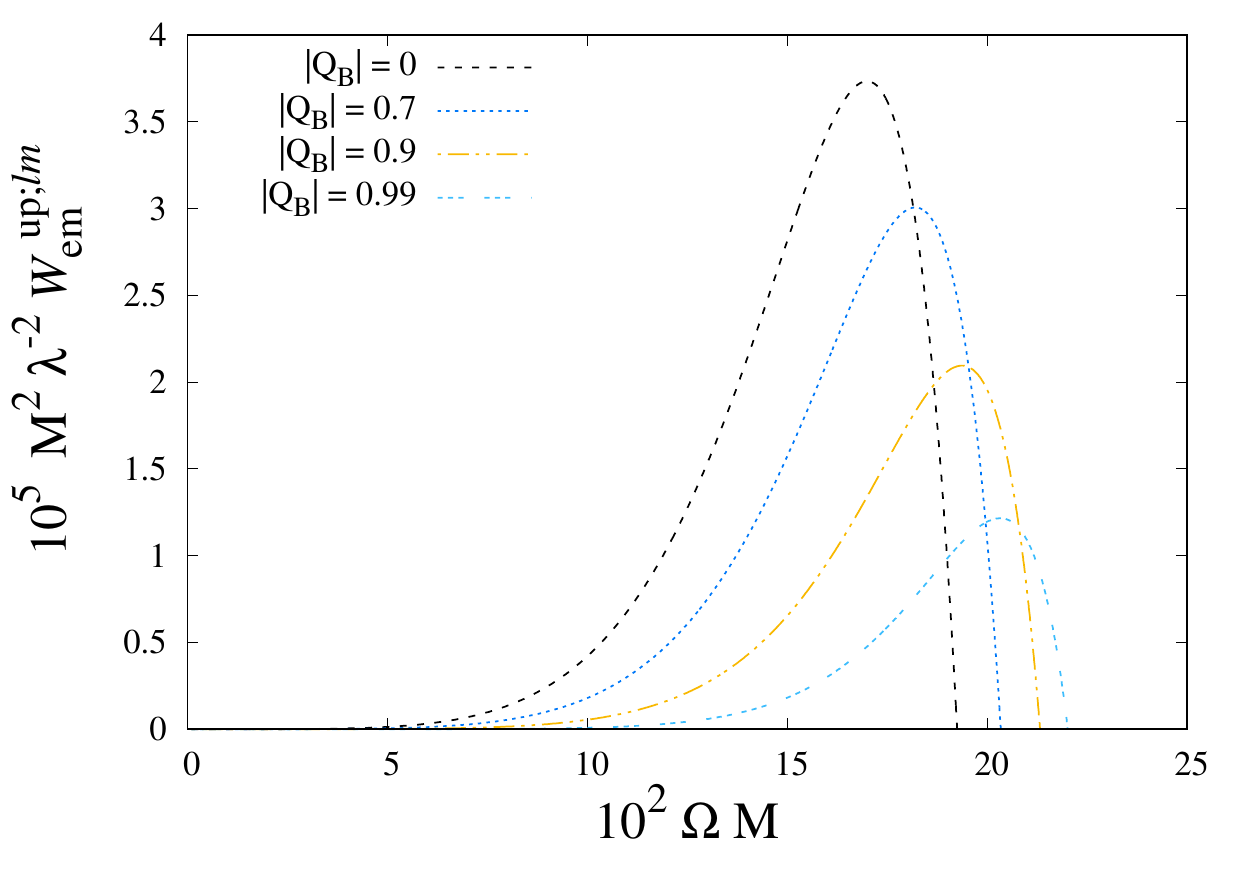} \includegraphics[scale=0.5,trim={1.9cm 0 0 0}, clip]{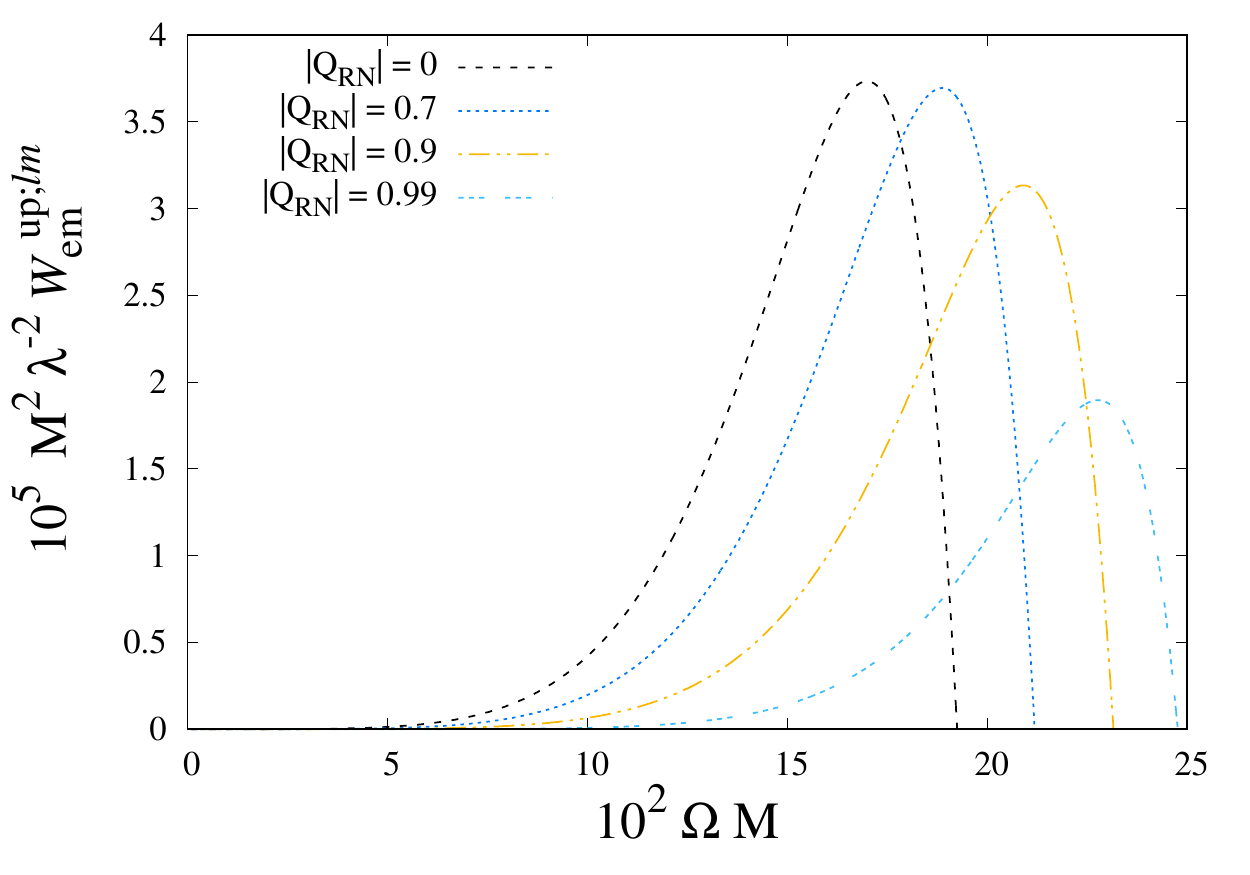}  \\
	\vspace{0.33cm}
	\includegraphics[scale=0.5]{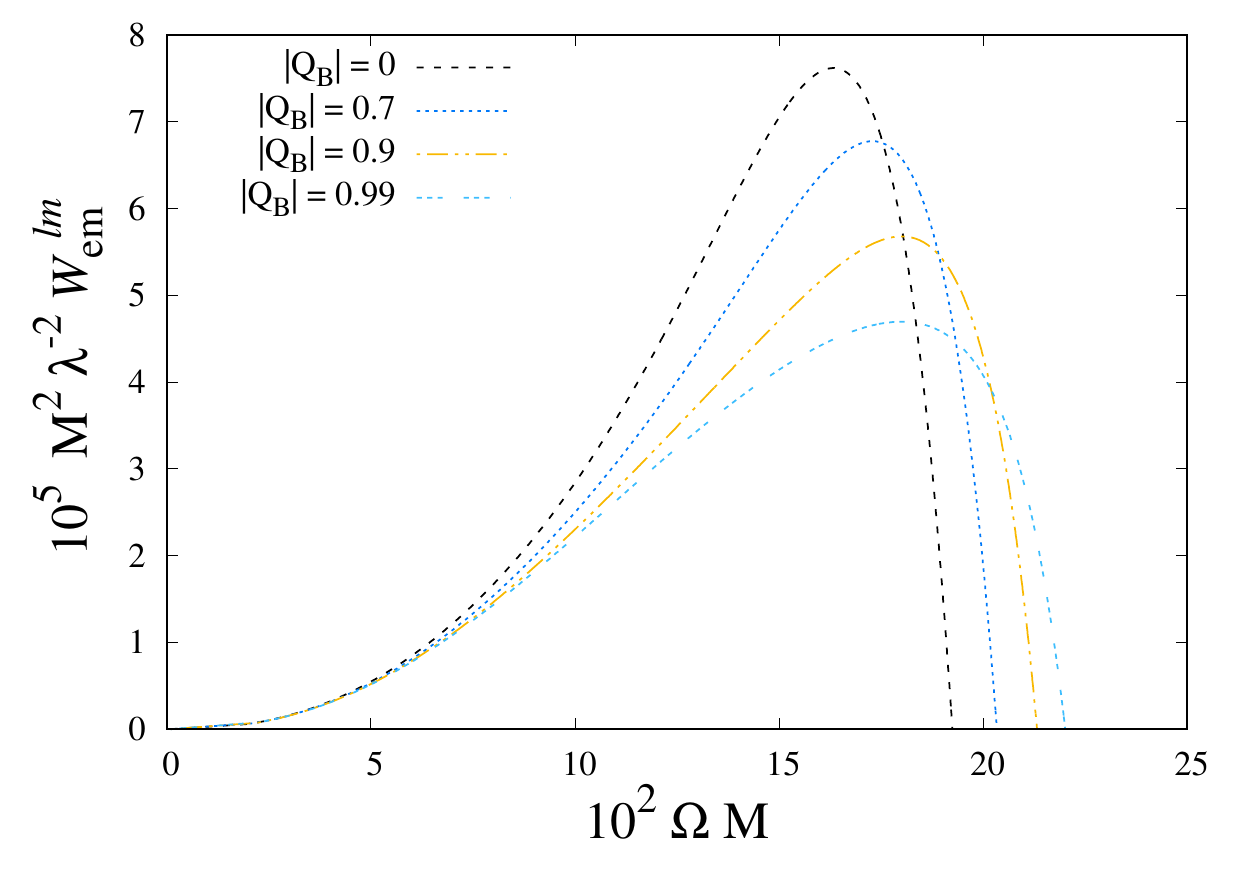} \includegraphics[scale=0.5,trim={1.7cm 0 0 0}, clip]{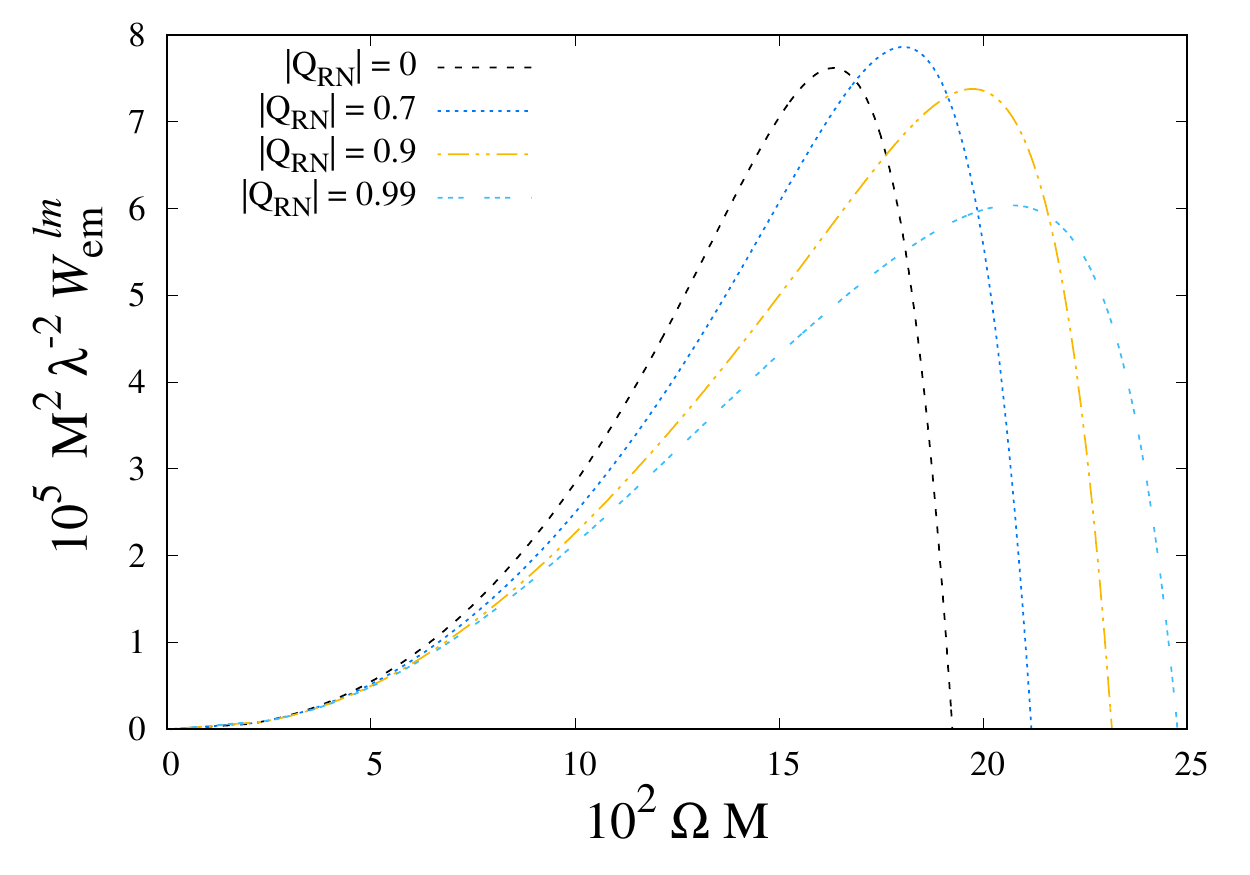}
	\caption{Emitted power for $in$ (\textit{top}), $up$ (\textit{middle}) modes and their sum (\textit{bottom}) of the $l=m=1$ mode, as a function of $\Omega$, in the Bardeen BH case (\textit{left}), and in the RN BH case (\textit{right}), with different values of $|Q_B|$ and $|Q_{RN}|$, respectively.}
	\label{fig:differentqs-total}
\end{figure*}

As shown in Fig.~\ref{fig:differentqs-total}, the peak of the total emitted power decreases as we increase the charge of the Bardeen BH. In the RN BH spacetime, we first see an increase in the peak height of the total emitted power as we increase the charge. From $Q_{RN} \gtrsim 0.7$, the behavior is similar to the Bardeen BH case.
As the charge in the RN case increases, the $in$ mode peaks display a significant increase in height (top plot in the right panel of Fig.~\ref{fig:differentqs-total}). Moreover, the \textit{up} mode peaks in the RN BH case fall off very slowly as the charge increases, as seen in the middle right panel of Fig.~\ref{fig:differentqs-total}.

\pagebreak
 
\pagebreak
 
\begin{figure}[h!]
	\centering
	\includegraphics[scale=1]{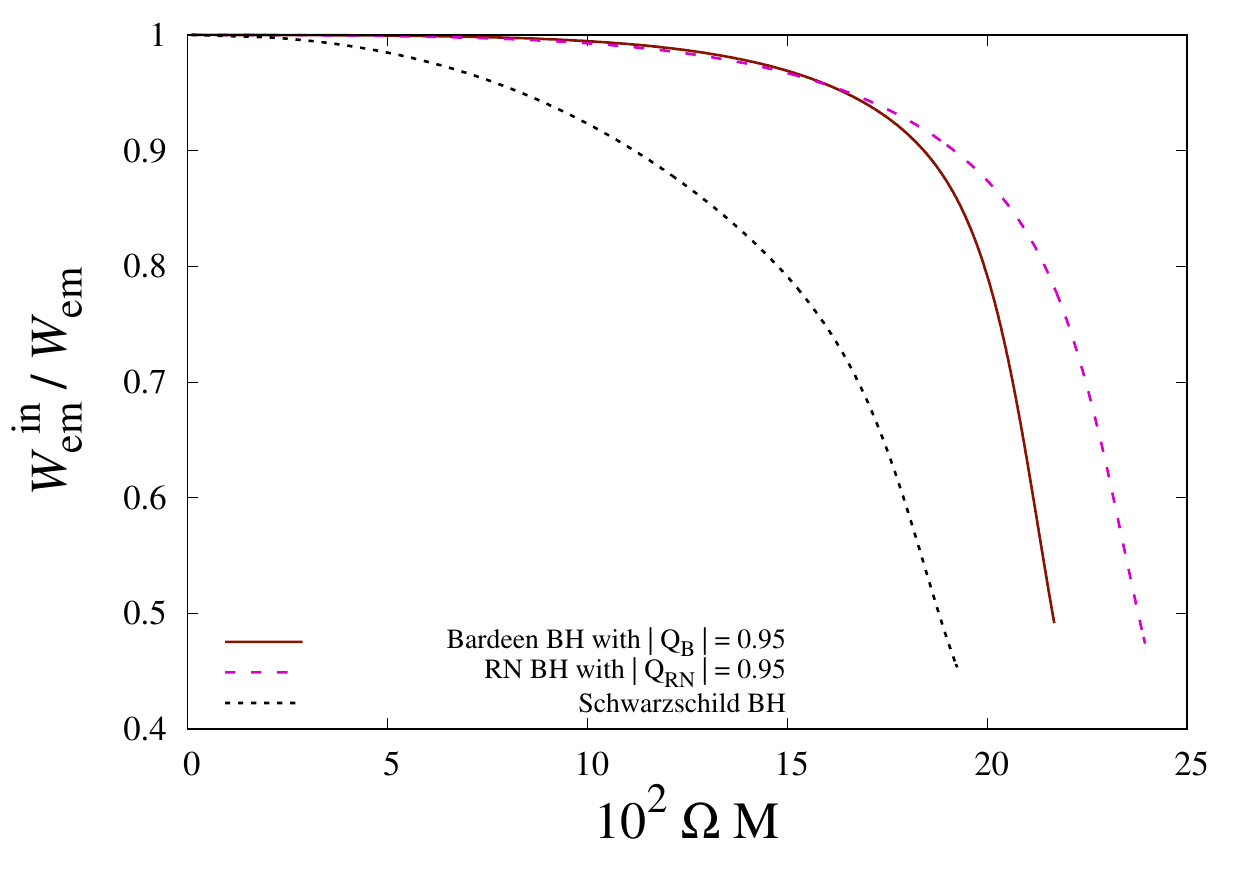}
	\caption{Ratio between the observed power at infinity, given by Eq.~(\ref{eq:observed-power}), and the total emitted power, given by Eq.~(\ref{eq:total-emitted-power}). The $l$ summations of Eqs.~(\ref{eq:total-emitted-power}) and (\ref{eq:observed-power}) is truncated at $l_{\mathrm{max}}=20$. The Bardeen BH and the RN BH have been chosen such that $|Q_{(i)}|=0.95$ in both cases.}
	\label{fig:ratio}
\end{figure}

We plot the ratio between the observed power at infinity, $W^{in}_{\mathrm{em}}$, given by Eq.~(\ref{eq:observed-power}), and the total emitted power, given by Eq.~(\ref{eq:total-emitted-power}) in Fig.~\ref{fig:ratio}. We see that, in both RN and Bardeen BH cases, they present a larger amount of radiation reaching null infinity, in comparison with the similar situation in a Schwarzschild BH. This is expected since their charge contribution to the metric has the sign opposite to the monopole (mass) term. The barrier in the effective potential, given by Eq.~(\ref{eq:effective-potential}), becomes smaller as we increase the absolute value of the charge (see Fig. \ref{fig:effectivepontential}), thus allowing more of the scalar radiation to reach null infinity.

\begin{figure}[h!]
	\centering
	\includegraphics[scale=1]{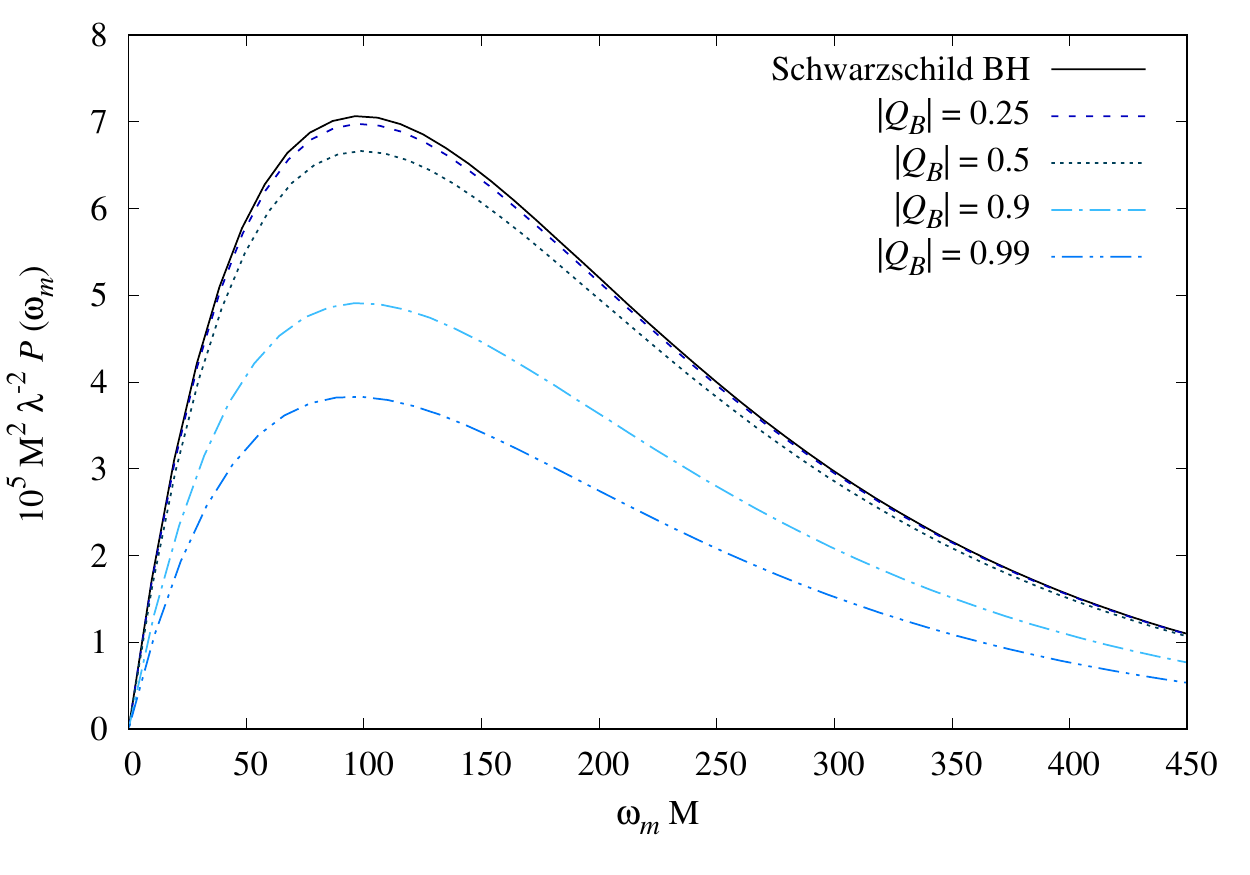}
	\caption{Spectral distributions for a Schwarzschild BH and Bardeen BHs with different values of the charge $|Q_{RN}|$. The source is located at $R/M=r_{\mathrm{ph}}/M+\delta$, with $\delta=10^{-4}$ and $r_{\mathrm{ph}}$ is the photon sphere radial position.}
	\label{fig:bardeen-spectrum}
\end{figure}

\begin{figure}[h!]
	\centering
	\includegraphics[scale=1]{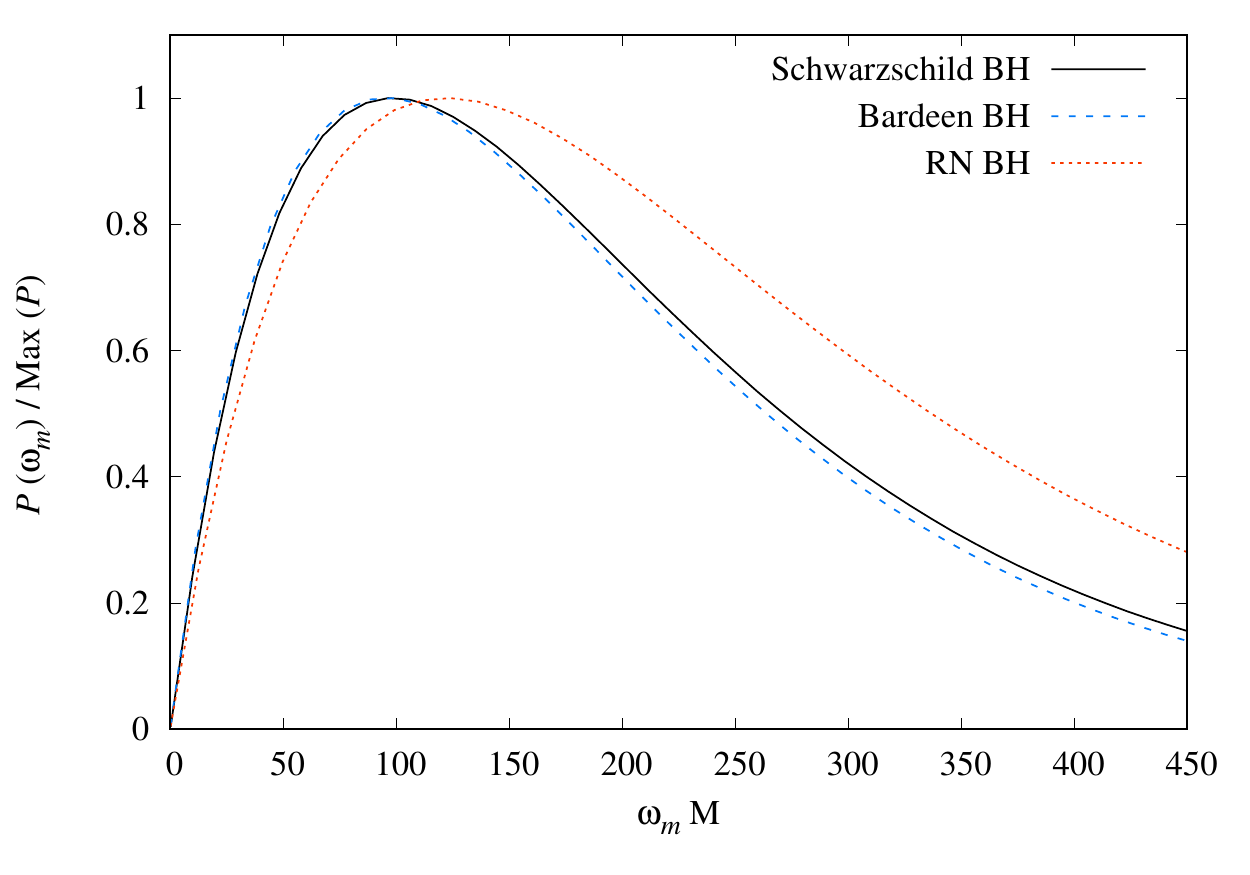}
	\caption{Normalized spectral distributions for three cases: a Schwarzschild BH, a Bardeen BH with $|Q_{B}|=0.99$ and a RN BH with charge $Q_{RN}=0.99$. The source is located at $R/M=r_{\mathrm{ph}}/M+\delta$, with $\delta=10^{-4}$ and $r_{\mathrm{ph}}$ is the photon sphere radial position. All the functions are normalized by their maximum achieved value.}
	\label{fig:normal-spectrum}
\end{figure}

For a given frequency $\omega_m=m\Omega$ of the emitted radiation observed at null infinity, we compute the spectral distribution by summing all the contributions from the different multipole modes to the total observed radiation, namely
\beq
P(\omega_m)=\sum_{l=m}^{\infty}W^{in;lm}_{\mathrm{em}}.
\eeq 
We compute the spectral disribution when the scalar source is located at $R/M=r_{\mathrm{ph}}/M+\delta$, where $r_{\mathrm{ph}}$ is the photon sphere radial position.\cite{Cardoso2009} We take $\delta=10^{-4}$, i.e. the source is very near the last unstable timelike circular orbit and the radiation exhibits a synchrotron behavior.\cite{Crispino2008} Although an emitting source in an unstable orbit would rapidly descend into the BH, this can be seen as a first approximation to the setting where the source is supported by an external force.
For the Bardeen BH case, the total amount of radiated power to infinity decreases (in comparison to the Schwarzschild case) as we increase the normalized charge, as we can see in Fig.~\ref{fig:bardeen-spectrum}. However, the spectral distribution displays a minor shift towards lower frequencies when we increase the charge, which can be noted in Fig.~\ref{fig:normal-spectrum}. 

\begin{figure}[h!]
	\centering
	\includegraphics[scale=1]{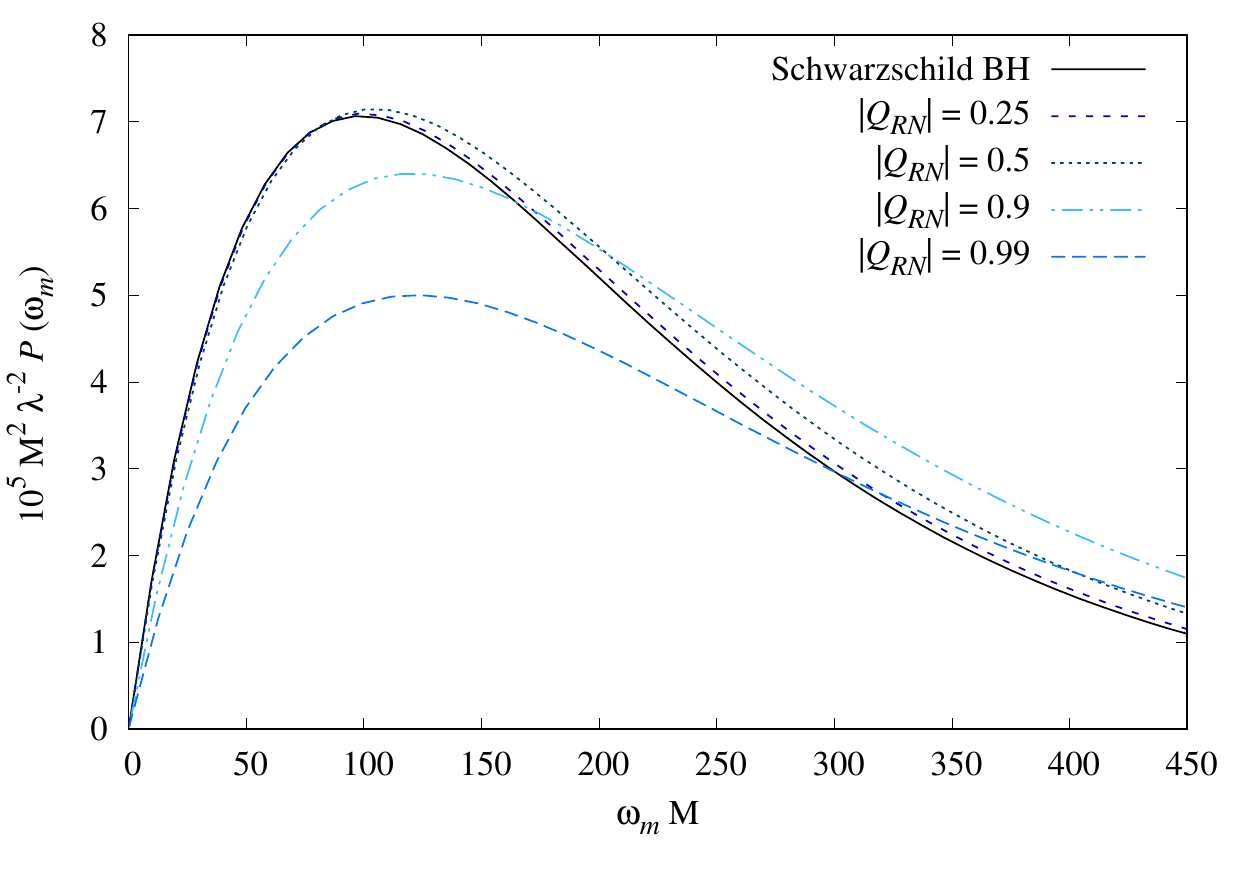}
	\caption{Spectral distributions for a Schwarzschild BH and RN BHs with different values of the charge $|Q_{RN}|$. The source is located at $R/M=r_{\mathrm{ph}}/M+\delta$, with $\delta=10^{-4}$, and $r_{\mathrm{ph}}$ is the photon sphere radial position.}
	\label{fig:RN-spectrum}
\end{figure}

In Fig.~\ref{fig:RN-spectrum}, we consider the RN BH case, which is more involved. At first, the total amount of radiation to infinity displays a slight increase when we increase the normalized charge. For the charge larger than a certain value, the value of the spectral distribution peak decreases. 

To better analyze the shifts in the spectral distribution in the three cases, we normalize them by their respective maximum values in Fig.~\ref{fig:normal-spectrum}, for representative values of the normalized charges. There is almost no change in the Bardeen BH case, besides a minor shift towards lower frequencies. In the RN BH case, there is a notable shift in the distribution towards higher frequencies. One can also note that, in all three cases, the spectral distribution peak is located at $m \approx 500$,\footnote{We recall that, for a source in circular orbit, the frequency $\omega$ is linked to the angular quantum number $m$ by Eq.~(\ref{eq:frequency-m}).} that is, most of the radiation is due to very high multipoles. 

\section{Final Remarks}
\label{sec:conclusion}

Regular black holes are remarkable solutions in General Relativity and other alternative theories of gravity and they provide a good platform to study the avoidance of singularities in geometric theories of gravity. Outside the horizon, Bardeen black holes are similar to a Reissner-Nordstr\"{o}m (or even a Schwarzschild) black hole. In this work, we have considered the radiation emission by a scalar source rotating around a general spherically symmetric black hole and analyzed some of its features when the black hole spacetime is the Schwarzschild, Reissner-Nordstr\"{o}m or Bardeen solution. In particular, we note their differences in the spectral distribution of the radiation, where it was shown to be very similar in the Bardeen and Schwarzschild cases. In comparison, the spectral distribution displays a shift towards higher frequencies in the Reissner-Nordstr\"{o}m solution setting.  

A possible extension of this work is to consider more general black holes, like rotating solutions, especially in regards to the frequency spectrum of the radiation and whether or not the source is co- or counterrotating with the black hole. Another important extension is to consider alternative theories of gravity, instead of exotic matter sources, in the radiation emission processes. In particular, an analysis of slowly rotating solutions in quadratic theories of gravity\cite{Stein2011,Pani2011} could provide us some insights and bounds to the deviations from GR. Finally, our analysis was done for emission of massless scalar particles. Although we expect a similar qualitative behavior for fields with different spins, like electromagnetic or gravitational radiation, the additional polarizations change the spectral distribution and it would be interesting to see if this same pattern occurs in the charged solutions considered here.

\section*{Acknowledgments}

We would like to thank A. Higuchi, L. C. B. Crispino and C. Macedo for useful discussions. This research was partially financed by Coordena\c{c}\~ao de Aperfei\c{c}oamento de Pessoal de N\'ivel Superior (CAPES, Brazil) -- Finance Code 001, and by Conselho Nacional de Desenvolvimento Cient\'ifico e Tecnol\'ogico (CNPq, Brazil). We also acknowledge funding from the European Union's Horizon 2020 research and innovation programme under the H2020-MSCA-RISE-2017 Grant No. FunFiCO-777740.


%
%
%
%
%
%


\end{document}